\documentclass[twocolumn,aps,superscriptaddress,amsmath,secnumarabic,pdflatex,amssymb]{revtex4-2}

\usepackage{graphicx}
\usepackage{bm}
\usepackage{amsmath,amsfonts,amssymb}
\usepackage{epstopdf}
\usepackage{xcolor}
\usepackage{color}
\usepackage{wasysym}

\usepackage{graphicx}
\usepackage{multirow}
\usepackage{dcolumn}

\begin{document}


\title{ Towards the demonstration of photon-photon collision with compact lasers }

\author{L. Q. Han}
\affiliation{Hunan Provincial Key Laboratory of High-Energy Scale Physics and Applications, School of Physics and Electronics, Hunan University, Changsha, 410082, China}

\author{J. Cai}
\affiliation{State Key Laboratory of Nuclear Physics and Technology, and Key Laboratory of HEDP of the Ministry of Education, CAPT, Peking University, Beijing, 100871, China}

\author{Y. R. Shou}
\affiliation{Center for Relativistic Laser Science, Institute for Basic Science, Gwangju, 61005, Republic of Korea}

\author{X. D. Liu}
\affiliation{Hunan Provincial Key Laboratory of High-Energy Scale Physics and Applications, School of Physics and Electronics, Hunan University, Changsha, 410082, China}

\author{J. Q. Yu}
\email{jinqing.yu@hnu.edu.cn}
\affiliation{Hunan Provincial Key Laboratory of High-Energy Scale Physics and Applications, School of Physics and Electronics, Hunan University, Changsha, 410082, China}

\author{X. Q. Yan}
\email{x.yan@pku.edu.cn}
\affiliation{State Key Laboratory of Nuclear Physics and Technology, and Key Laboratory of HEDP of the Ministry of Education, CAPT, Peking University, Beijing, 100871, China}
\affiliation{Collaborative Innovation Center of Extreme Optics, Shanxi University, Taiyuan, Shanxi, 030006, China.}
\affiliation{Guangdong Laser Plasma Institute, Guangzhou, 510540, China.}


\begin{abstract} 
We report a proposal to observe the two-photon Breit-Wheeler process in plasma driven by compact lasers. A high charge electron bunch can be generated from laser plasma wakefield acceleration when a tightly focused laser pulse transports in a sub-critical density plasma. The electron bunch scatters with the laser pulse coming from the opposite direction and results the emitting of high brilliance X-ray pulses. In a three-dimensional particle-in-cell simulation with a laser pulse of $\sim$10 J, one could produce a X-ray pulse with photon number higher than $3\times10^{11}$ and brilliance above $1.6\times 10^{23}$ photons/s/mm$^2$/mrad$^2$/0.1$\%$BW at 1 MeV. The X-ray pulses collide in the plasma and create more than $1.1\times 10^5$ electron-positron pairs per shot. It is also found that the positrons can be accelerated transversely by a transverse electric field generated in the plasma, which enables the safe detection in the direction away from the laser pulses. This proposal which has solved key challenges in laser driven photon-photon collision could demonstrate the two-photon Breit-Wheeler process on a much more compact device in a single shot.
\end{abstract}

\maketitle

\section{Introduction}

The Breit-Wheeler (BW) process, which was first proposed in theory by Breit and Wheeler in 1934 \cite{breit1934}, could create matter and antimatter from the collision of two photon quanta. Normally, the BW process includes two-photon process ($\gamma + \gamma  \to e^- + e^+$) and multi-photon process ($\gamma + n\gamma \to e^- + e^+$). The BW process is recognized as a fundamental component of quantum electrodynamics theories \cite{marklund2006}. Hence, the experimental study of the BW process could provide important tests for the fundamental physics. Nevertheless, due to the lack of high brilliance and energetic photon sources, Breit and Wheeler even declared in 1934 that it was hopeless to observe pair creation in the laboratory. In the past forty years, the CPA technology \cite{strickland1985} has greatly promoted the development of high-power laser technology \cite{danson2019}, and the laser intensity above $10^{23}$ W/cm$^2$ has been realized most recently \cite{yoon2021}. At such intensity, it is very likely to emit high energy photons above GeV \cite{ridgers2012}. The multi-photon BW process has been experimentally demonstrated in 1997 through the scattering between high energy photon of GeV and several laser photons \cite{burke1997}. In the past decade, increasing attention has been paid to experimentally study the multi-photon BW process with high-power PW lasers \cite{hu2010,ridgers2012}. Numerical simulations have shown that one could generate $10^{11}$ electron-positron pairs per shot by using 10-PW lasers \cite{zhu2016}.

On the other hand, the two-photon BW process involving real photon, which is a fundamental phenomenon in quantum electrodynamics theories, has never been observed in the laboratory. The difficulty in experimentally observing the two-photon BW process lies in the absence of high brilliance $\gamma$-ray pulses \cite{golub2021} together with the challenge in extracting useful signals from noise \cite{pike2014,ribeyre2016,yu2019}. In the past few years, various theoretical scenarios have been proposed to observe the signal of two-photon BW process. Pike $et$ $al.$ presented the first photon-photon collider in a vacuum hohlraum \cite{pike2014}. Ribeyre $et$ $al.$ proposed an experimental approach using MeV photon sources driven by $\sim$10 PW lasers \cite{ribeyre2016}. Micieli $et$ $al.$ designed a photon-photon collider based on conventional Thomson $\gamma$-ray sources \cite{micieli2016}. Yu $et$ $al.$ put forward an approach utilizing collimated and wide-bandwidth $\gamma$-ray pulses \cite{yu2019}. Wang $et$ $al.$ systematically studied the two-photon pair production using a structured target in three-dimensional Particle-in-cell simulations \cite{wang2020}. He $et$ $al.$ discussed the dominance condition of $\gamma$-$\gamma$ collision in plasma driven by high-intensity lasers \cite{he2021}. Zhao $et$ $al.$ investigated the polarization characteristics of linear BW in polarized $\gamma$-$\gamma$ collider \cite{zhao2022}. In brief, the above solutions require a large machine like a laser near 10 PW or a conventional accelerator. However, these large machines demand considerable financial investments and extensive space to accommodate them. Therefore, the more interesting thing is realizing this scientific experiment on a compact device.

Most recently, a scientific team led by the UK is attempting to conduct the experiment of photon-photon collision on a compact platform \cite{kettle2021}. Due to the generation of only around $7\times 10^7$ photons per shot, more than $10^9$ shots are needed to observe the linear BW process on this platform, and it could be reduced to 5000 shots by increasing the electron energy. However, the current technology still poses a challenge to ensure the stability of the laser over such a large number of shots. Hence, further developments of such a compact platform should significantly reduce the number of shots to a few hundred or even to a single one, making this process more achievable.

In this paper, we propose a scheme to observe the two-photon Breit-Wheeler process in a single shot through the interaction between a sub-critical density plasma (SCDP) and compact lasers. The laser pulses were tightly focused to a spot of several micro-meters so that they can achieve an intensity higher than 10$^{21}$ W/cm$^2$ despite the relatively low power of hundreds of TW, which can be easily achieved on a compact platform. When the laser pulses propagate through the SCDP, laser plasma wakefield acceleration can be aroused and high-charge electron bunches would be generated. Thomson scattering \cite{thomson1883,yan2017} occurs when the electron bunch scatters with the laser pulse coming from the opposite direction, resulting in the emission of high brilliance X-ray pulses. Three-dimensional particle-in-cell (3D-PIC) simulation with laser pulses of $\sim$10 J and a SCDP of $0.1n_c$ demonstrates that it is possible to generate high-charge electron bunches of $\sim$10 nC, X-ray pulses with photon number higher than $3\times 10^{11}$ and brilliance above $1.6\times 10^{23}$ photons/s/mm$^2$/mrad$^2$/0.1$\%$BW at 1 MeV. Shortly after the Thomson scattering, the X-ray pulses collide in a ultra small transverse area and create more than $1\times 10^5$ electron-positron pairs in a single shot. It is also found that the positrons can be deflected by a transverse electric field generated in the plasma, which enables the safe detection in the direction away from the laser pulses. Therefore, the observation of two-photon Breit-Wheeler process can be realized on a much more compact device in a single shot.

\section{ Scheme design and simulation setup }

With the development of target manufacturing technology, it has become possible to produce SCDP in the laboratory using carbon nanotube \cite{ma2007}. In the past few years, SCDP has been widely used to produce high-quality charged particles \cite{liu2013,hu2015,ma2019} and radiation \cite{lobok2018,liu2019,shen2021,shou2023}. Recently, we reported a regime in which a high-charge electron bunch was generated through a combined process of laser plasma wakefield acceleration (LPWA) and direct laser acceleration in a SCDP, driven by a tightly focused laser pulse \cite{huang2023}. In this regime, an electron bunch with a charge higher than 10 nC could be generated using a compact laser.

\begin{figure}[htb]
    \centering
    \includegraphics[width=1\linewidth]{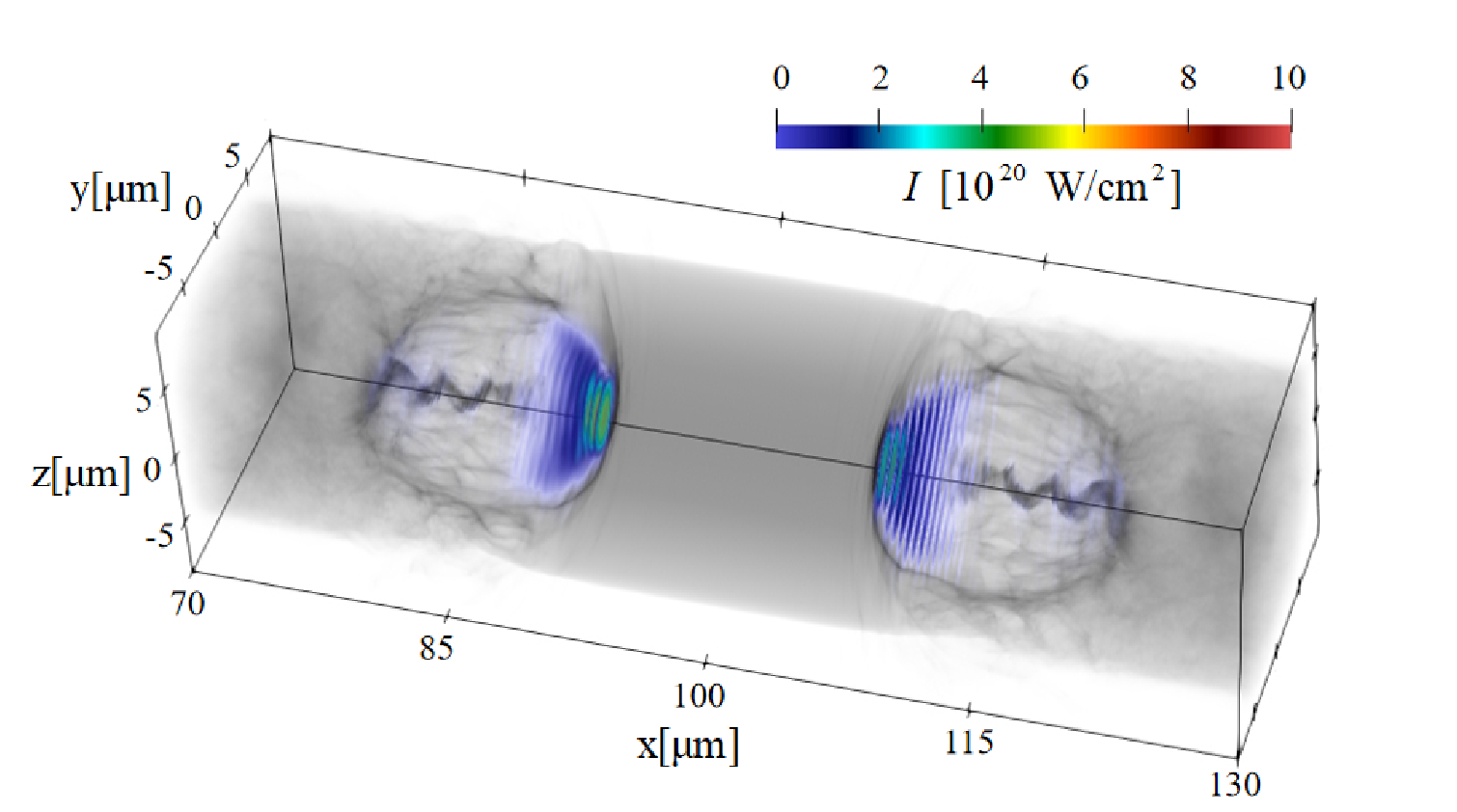}
    \caption{ Schematics for the setup of photon-photon collision in plasma driven by compact lasers of $\sim$10 J. The laser pulses, which are tightly focused to an intensity above 10$^{21}$ W/cm$^2$, arouse wake-field acceleration for the generation of high charge electron bunches ($\sim$10 nC) in the sub-critical density plasma of $\sim$200 $\mu$m. There are two stages that can generate high brilliance $\gamma$-ray pulses. Betatron X-ray can be radiated when the electron bunches oscillate transversely in the plasma bubble during the stage of electron acceleration. Thomson scattering occurs when the electron bunches collide with the laser pulse coming from the opposite direction. The high brilliance $\gamma$-ray pulses collide in the plasma and create more than $1\times 10^5$ electron-positron pairs. }
    \label{fig1}
\end{figure}

Following the route of generating a high-charge electron bunch, we propose a scheme to observe photon-photon collision in plasma driven by compact ultra-short pulse lasers as shown in Fig. \ref{fig1}. Two laser pulses with the energy of $\sim$10 J are tightly focused to a spot of several micro-meters. The laser pulses which irradiate on a $\sim$200 $\mu$m SCDP from the opposite directions result in the generation of high charge electron bunches dominated by laser plasma wakefield acceleration \cite{huang2023}. When the electron bunch collides with the laser pulse from the opposite direction, high-order Thomson scattering \cite{thomson1883,yan2017} will take place, inducing the radiation of high brilliance $\gamma$-ray pulse. In this setup, the laser pulse should collide with the electron bunch before being fully depleted. Since the electron bunch is closely behind the driving laser, one could achieve photon-photon collision in a spot of several micro-meters shortly after the generation of the $\gamma$-ray pulses.

We used 3D-PIC code Epoch3d \cite{arber2015} to explore the generations of the high-charge electron bunches and $\gamma$-ray pulses, as well as the dynamic behavior of positrons in the plasma. The simulation box X $\times$ Y $\times$ Z  = 200 $\mu$m $\times$ 24 $\mu$m $\times$ 24 $\mu$m was divided into $4000\times 240 \times 240$ cells. The driving lasers, each producing a pulse of 21.3 fs duration at FWHM (full width at half maximum), operated at a wavelength $\lambda _0$ = 800 nm and had a Gaussian spatial profile, a $sin^2$ temporal profile, and a Z-axis polarization. In case A, these pulses were tightly focused to a waist $w_0$ = 2.4 $\mu$m with a normalized intensity $a_0$ = 40. The on target power of each pulse was about 10 J which could be exported by a compact laser system \cite{danson2019}. Under the laser pulses of ultra high intensity, the SCDP could be fully ionized to electrons and C$^{6+}$. In the simulations, a cylindrical SCDP was modelled with an electron density of $n_e = 0.1n_c$, a length of 196 $\mu$m and a radius of 12 $\mu$m. Here, $n_c$ is the critical density of the plasma. 20 (4) macro-electrons (C$^{6+}$) were initialized into each cell.

\section{ Electron acceleration and $\gamma$-ray generation }

\begin{figure}[htb]
    \centering
    \includegraphics[width=1\linewidth]{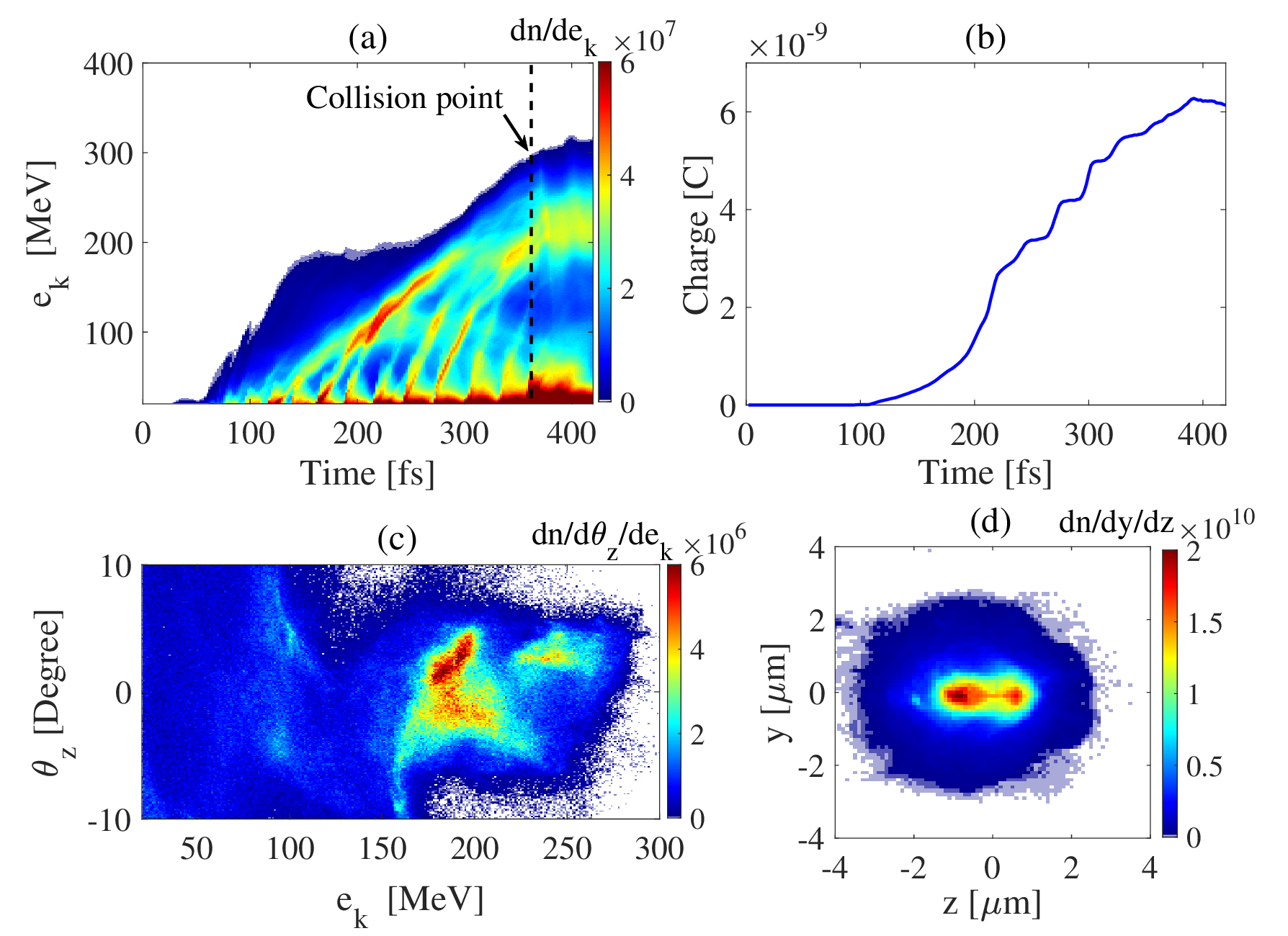}
    \caption{ 3D-PIC simulation results of laser wakefield acceleration in the 200 $\mu$m SCDP driven by the laser pulses of case A. (a) Evolution of the electron kinetic energy spectra with simulation time to 420 fs. (b) The charge of a single electron bunch ($>$ 100 MeV) as a function of the simulation time. (c) The spectral-angular distribution and (d) the spatial distribution of the electrons at 350 fs.  }
    \label{fig2}
\end{figure}

Figure \ref{fig1} shows the distributions of laser intensity and electron density driven by the laser pulses of case A at 330 fs, from which one can see two plasma cavities and the accelerated electron bunches in the wakefield. As demonstrated in the previous work \cite{huang2023}, LPWA could dominate the electron acceleration in the SCDP when the normalized laser intensity $a_0$ is larger than 20. Fig. \ref{fig2}(a) illustrates the evolution of the electron energy spectra in the first 420 fs. It demonstrates that the electrons can be accelerated to $\sim$300 MeV in a distance less than 100 $\mu$m, corresponding to an acceleration gradient of roughly 3 TV/m. The figure clearly shows a continuous capture of electrons and indicates that the beam-loading effect \cite{rechatin2009} is no longer working under this strong nonlinear condition, leading to the charge of the electron bunch exceeding 6 nC at 420 fs as shown in Fig. \ref{fig2}(b). Since most of the laser has not been depleted yet, more electrons could be captured by the wakefield and accelerated to higher energy than that shown in Fig. \ref{fig2}(b) in the case of a longer SCDP.

The production of electron-positron pairs is a function of the photon number, the collision area size and the photon-photon cross-section for BW. Comparing with the normal LPWA \cite{lu2007,esarey2009}, we pursued high-charge and collimated electron bunches at the cost of their energy spread and divergence angle in this work. High-charge electron bunches could radiate a high flux photon pulse which is a prerequisite for enhancing the electron-positron pair production. Meanwhile, the impacts of energy spread and divergence angle on the pair production is virtually negligible in this proposal. Figure \ref{fig2}(c) shows the spectral-angular distribution of the electron bunch before the collision, the divergence angle in the polarized direction $\theta_z \sim$5$^\circ$ ($\theta_y \sim$2$^\circ$) is larger than that in normal LPWA \cite{lu2007,esarey2009}. The divergence angle of the electron bunch could be further reduced assuming a longer acceleration distance. The transverse area of the electron bunch is smaller than the spot of the laser from the opposite direction, as shown in Fig. \ref{fig2}(d), making sure the enough overlapping space during collision for Thomson scattering.

\begin{figure}[htb]
    \centering
    \includegraphics[width=1\linewidth]{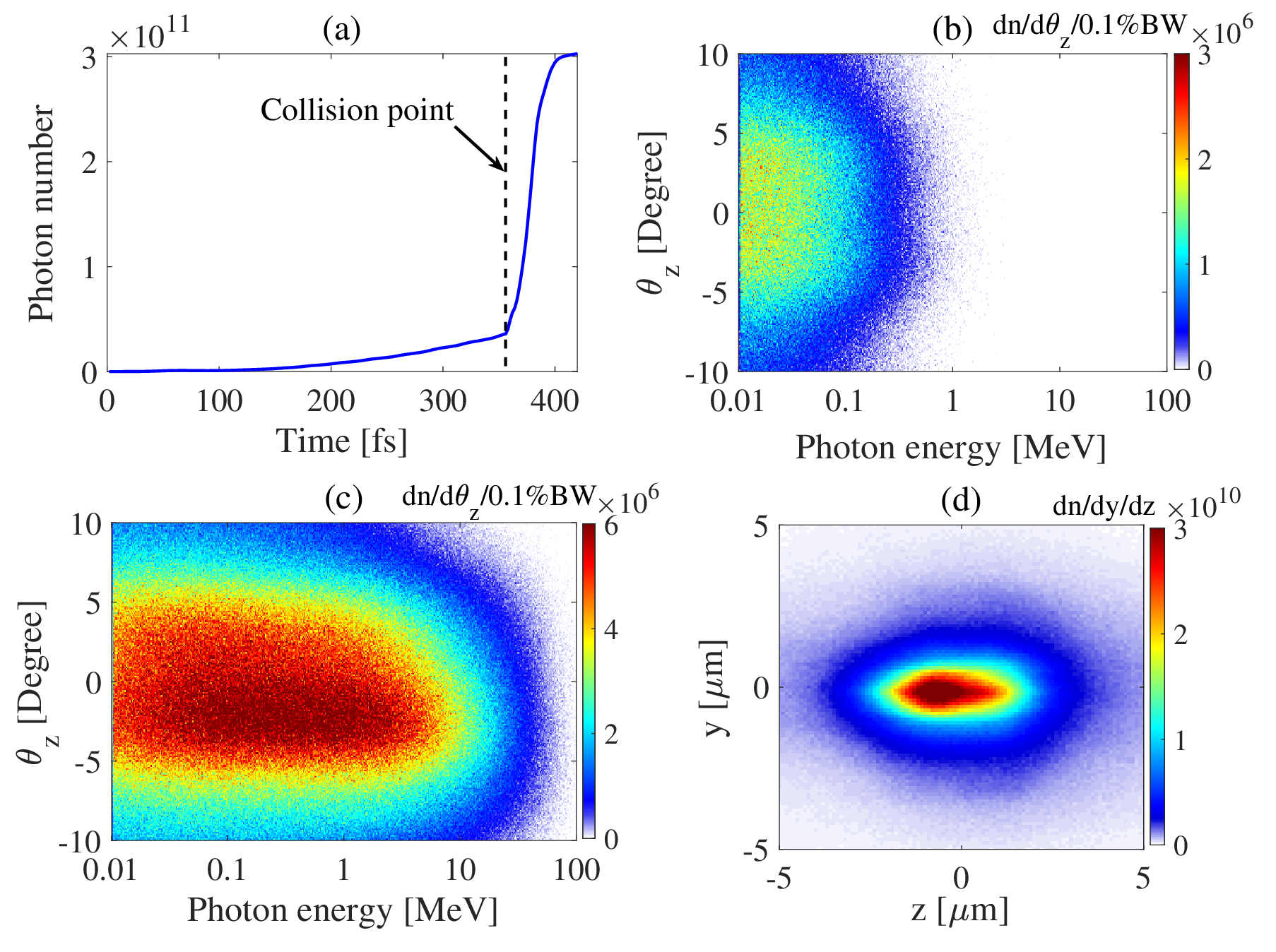}
    \caption{ 3D-PIC simulation results of X-ray photons ($>$ 10 keV) generated from the 200 $\mu$m SCDP driven by the laser pulses of case A. (a) The photon number as a function of the simulation time. The spectral-angular distribution of the X-ray photons (b) at 350 fs and (c) 420 fs. (d) The spatial distribution of the X-ray photons at 420 fs.}
    \label{fig3}
\end{figure}

In the simulation, the electron bunch and the laser pulse collided at 358 fs. Before the collision, Betatron radiation \cite{rousse2004,kiselev2004} dominates the generation of the X-ray photons. There were $3.6\times 10^{10}$ X-ray photons ($>$ 10 keV) in the simulation box at 350 fs. This number is smaller than the actual generated number since the photons with large divergence angle had already moved out of the simulation box and failed to be counted. From the spectral-angular distribution of the X-ray photons shown in Fig. \ref{fig3}(b), one can see that most of the photons have the energy less than 100 keV, and the divergence angle is $\theta_z \sim$7.2$^\circ$. The Thomson scattering takes place from 358 to 400 fs, resulting in the generation of more than $2.65\times 10^{11}$ X-ray photons ($>$ 10 keV) during the collision as displayed in Fig. \ref{fig3}(a). The photon generation duration time suggests that only a small portion of the laser pulse contributed to electron acceleration.

Figure \ref{fig3}(c) shows the spectral-angular distribution of the X-ray photons after collision. It is found that the photon maximum energy has been significantly enhanced and the divergence angle was reduced to $\theta_z \sim$6.3$^\circ$. Since a large number of photons had energy higher than 0.511 MeV, which is the threshold energy for electron-positron pair generation in photon-photon collision \cite{breit1934,burke1997}, the generation of electron-positron pair through the two-photon BW process becomes possible. In the photon-photon collision, one could significantly increase the number of electron-positron pairs by making the photon-photon collision in a smaller transverse area. The spatial distribution of the photons shown in Fig. \ref{fig3}(d) indicates that the transverse area of the photon bunch is 3.8 $\mu$m$^2$. Then we can calculate the brilliance at 1 MeV, yielding a value of greater than $1.6\times 10^{23}$ photons/s/mm$^2$/mrad$^2$/0.1$\%$BW. Despite the lack of a significant advantage in brightness, this photon bunch has a much smaller light spot compared to the previous Thomson light source with SCDP \cite{liu2019}. This characteristic is more advantageous for photon-photon collision.

\section{ Photon-photon collision }

\begin{figure}[htb]
    \centering
    \includegraphics[width=1\linewidth]{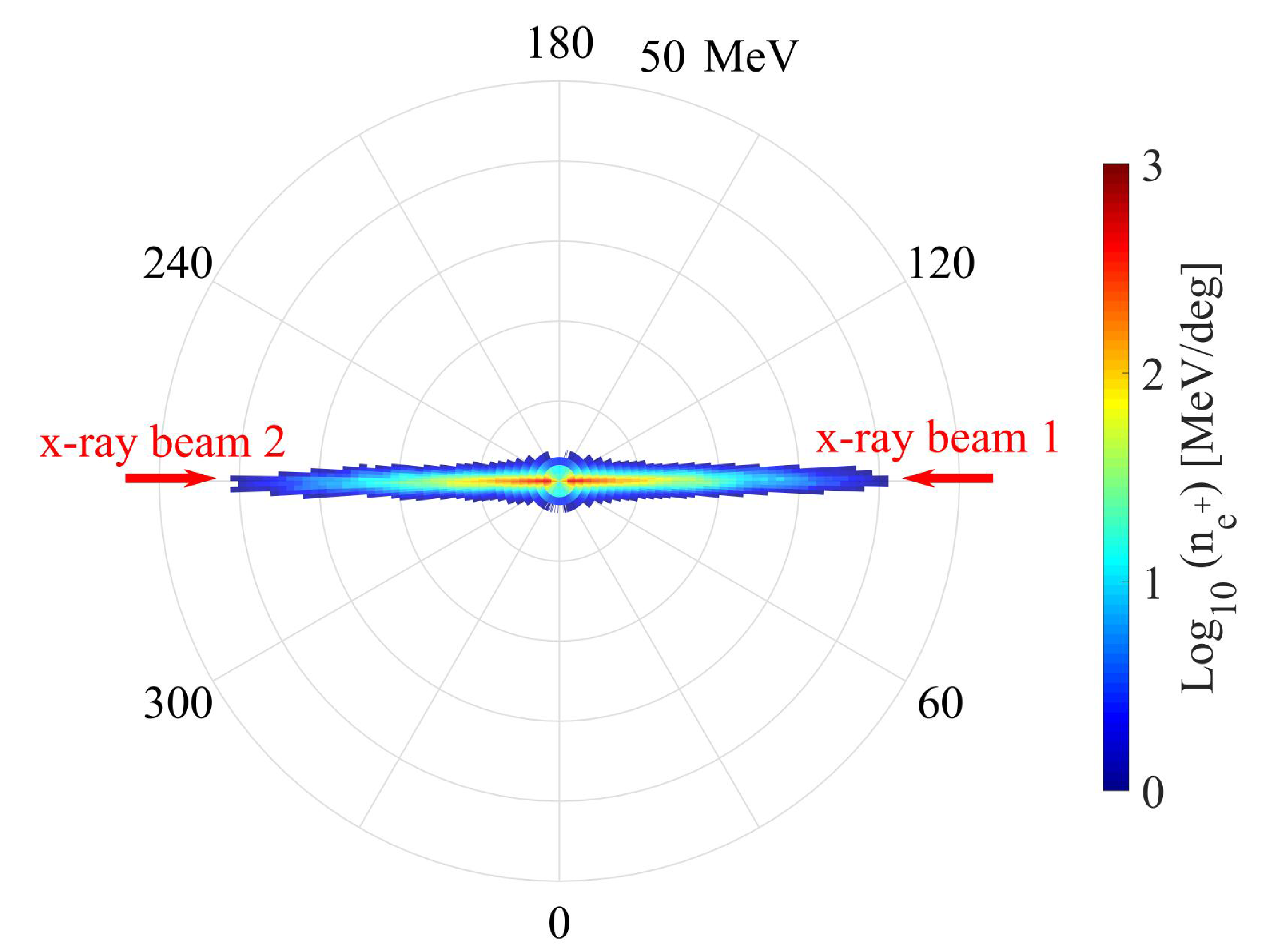}
    \caption{ The angular-spectrum distribution of the positrons calculated from the photon-photon collision in the 200 $\mu$m SCDP driven by the laser pulse of case A. }
    \label{fig4}
\end{figure}

After obtaining the spectral-angular distribution of the X-ray photon shown in Fig. \ref{fig3}(c), we adopt the method of macro-photon \cite{yu2019} to calculate the electron-positron pair creation. In the photon-photon collision for electron-positron pair creation, the threshold condition is  
\begin{equation}
 S = \frac {\varepsilon_{\gamma_1} \varepsilon_{\gamma_2} ( 1 - \cos\theta_c)}{2(m_ec^2) ^2} \ge 1,
\label{equ1}
\end{equation}

where $\theta_c$ is the collision angle, $ \varepsilon_{\gamma_1}$ and $ \varepsilon_{\gamma_2}$ are the average energies of the macro-photons, $m_e$ is the rest mass of the electron, $c$ is the vacuum light speed. In each collision between the macro-photons above the threshold condition, a macro-pair will be generated. The weight value of the macro-pair can be expressed as 
\begin{equation}
 w_{pair} = \frac { w_{\gamma_1} w_{\gamma_2} \sigma_{\gamma_1 \gamma_2}} { S_c },
 \label{equ2}
\end{equation} 
where $w_{\gamma_1}$ and $w_{\gamma_2}$ are the weight value of the macro-photons, $S_c$ is the transverse size of cross area of the photon bunches, the cross-section $\sigma_{\gamma_1 \gamma_2}$ of the BW process is  
\begin{equation}
 \sigma_{\gamma_1 \gamma_2} = \frac {\pi}{2} r_0^2 (1-\beta^2) [(3-\beta^4) \ln(\frac{1+\beta}{1-\beta}) - 2\beta(2-\beta^2)    ] .
 \label{equ3}
\end{equation} 
Here $\beta = (1-1/S)^{1/2}$ and $r_0$ is the electron classical radius. Then, one needs to get the energy $\varepsilon_e$ ( $\varepsilon_p$) and momentum $\vec{p_e}$ ($\vec{p_p}$) of the electron (positron) in the laboratory frame \cite{yu2019}.

In this proposal, the positrons can be created through the photon-photon collision in the SCDP driven by the laser pulses of case A as shown in Fig. \ref{fig1}. From the simulation, it is found that the effective photon-photon collision time is from $\sim$370 fs to $\sim$410 fs, and the effective transverse area for collision $S_c$  is less than 3.0 $\mu$m$^2$. Then we used X-ray pulses with similar parameters shown in Fig. \ref{fig3} to calculate the creation of the electron-positron pair. In the calculation, $1.1\times 10^5$ electron-positron pairs could be created and the angular-spectrum distribution of the positrons was plotted in Fig. \ref{fig4}. Since wide bandwidth and collimated X-ray pulses \cite{yu2019} were used here, the created positrons had a small divergence angle of 4$^\circ$.

\begin{figure}[htb]
    \centering
    \includegraphics[width=1\linewidth]{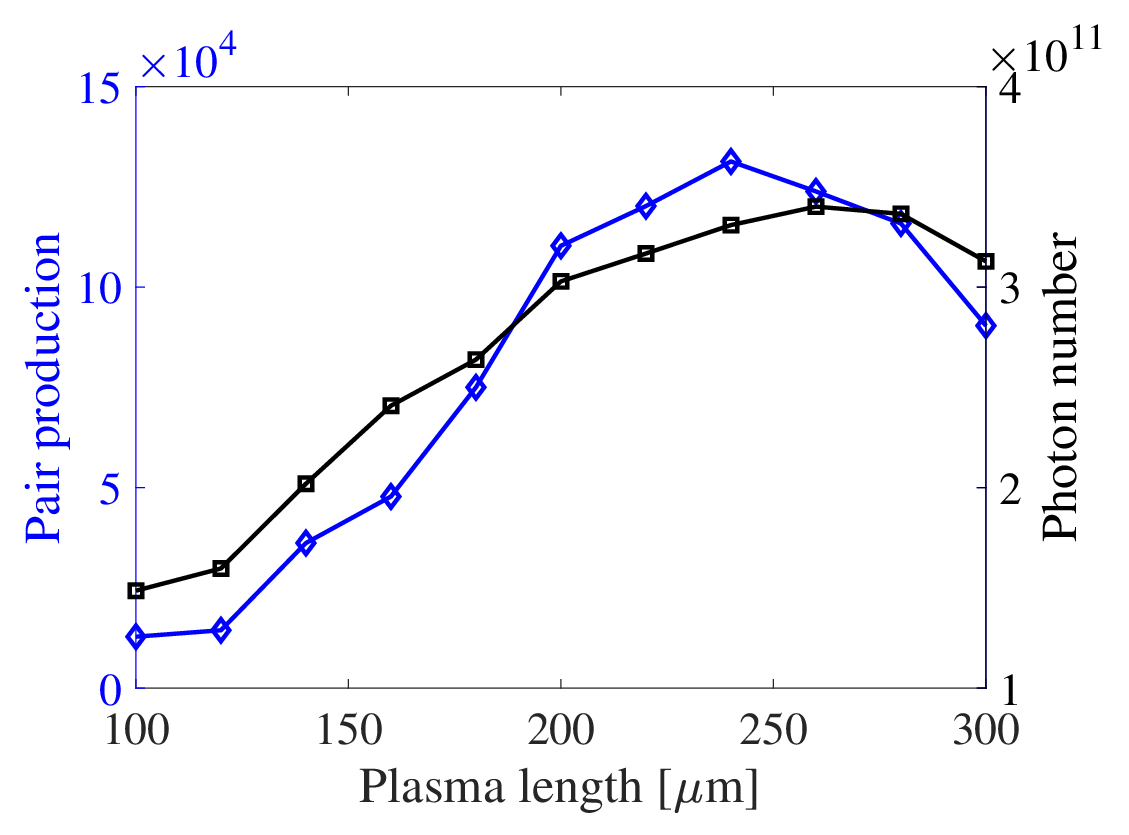}
    \caption{The effect of the plasma length on X-ray photon number and pair production by varying the plasma length from 100 to 300 $\mu$m driven by the laser pulse of case A.}
    \label{fig5}
\end{figure}

From the evolution of the electron kinetic energy spectra shown in Fig. \ref{fig2}(a), one can see that the plasma length significantly affects the results of electron acceleration. In the case of a short length plasma, the electrons cannot be effectively accelerated, leading to a substantial reduction in both photon number and pair number. For longer plasma, most of the laser pulses has been consumed before the Thomson scattering, leaving a very small portion of the laser pulses for scattering. Hence, there should be an optimal plasma length for the generation of X-ray bunch. In the simulations, the plasma length was varied from 100 $\mu$m to 300 $\mu$m to examine the impact on photon number and pair number. As shown in Fig. \ref{fig5}, the photon number increased with the plasma length from $1.5\times$10$^{11}$ to a peak of $3.4\times$10$^{11}$ at a plasma length of 260 $\mu$m, then decreased to $3.1\times$10$^{11}$ at 300 $\mu$m. Simultaneously, the pair number also experienced a rise from $1.3\times$10$^{4}$ to a peak of $1.3\times$10$^{5}$ at 240 $\mu$m, before declaring to $9\times$10$^{4}$ at 300 $\mu$m. Therefore, we can infer that the optimal plasma length interval is between 240 $\mu$m and 260 $\mu$m, for the laser pulse of case A.

\begin{table}[htb]
    \centering
    \renewcommand\arraystretch{1.5}
    \begin{tabular}{|c|c|c|}
    \hline  \textbf{ Laser $a_0$ } & \textbf{ Photon number } & \textbf{ Pair number } \\
    \hline  25  & $6.2\times$10$^{10}$  &  $2.5\times$10$^3$ \\ 
    \hline  30  &  $1.8\times$10$^{11}$  &  $5.7\times$10$^4$ \\  
    \hline  40  &  $3.0\times$10$^{11}$  &  $1.1\times$10$^5$ \\ 
    \hline  60  &  $9.8\times$10$^{11}$  &  $1.2\times$10$^6$ \\
    \hline  80  &  $1.8\times$10$^{12}$  &  $5.0\times$10$^6$ \\  
    \hline
    \end{tabular}
    \caption{The effect of the laser normalized intensity $a_0$ on X-ray photon number and pair production by fixing the SCDP to 200 $\mu$m and laser pulse waist $w_0$ to 2.4 $\mu$m.}
    \label{tab:my_label}
\end{table}

In addition, a series of 3D simulations were performed to investigate the influence of the laser normalized intensity $a_0$ on X-ray photon number and pair number. In the simulations, $a_0$ was varied from 25 to 80, while the SCDP length and the laser pulse waist were fixed to 200 $\mu$m and 2.4 $\mu$m respectively. From the simulation results listed in table \ref{tab:my_label}, it is found that the photon number is proportional to $a_0^{2.4}$, and the pair number is proportional to $a_0^{5.0}$. The the scaling law indicates that it is easy to create 10$^6$ pairs with 1-PW laser pulses, and could create 10$^9$ pairs by using 10-PW laser pulses.

\section{ Dynamics of the positrons in SCDP }

\begin{figure}[htb]
    \centering
    \includegraphics[width=1\linewidth]{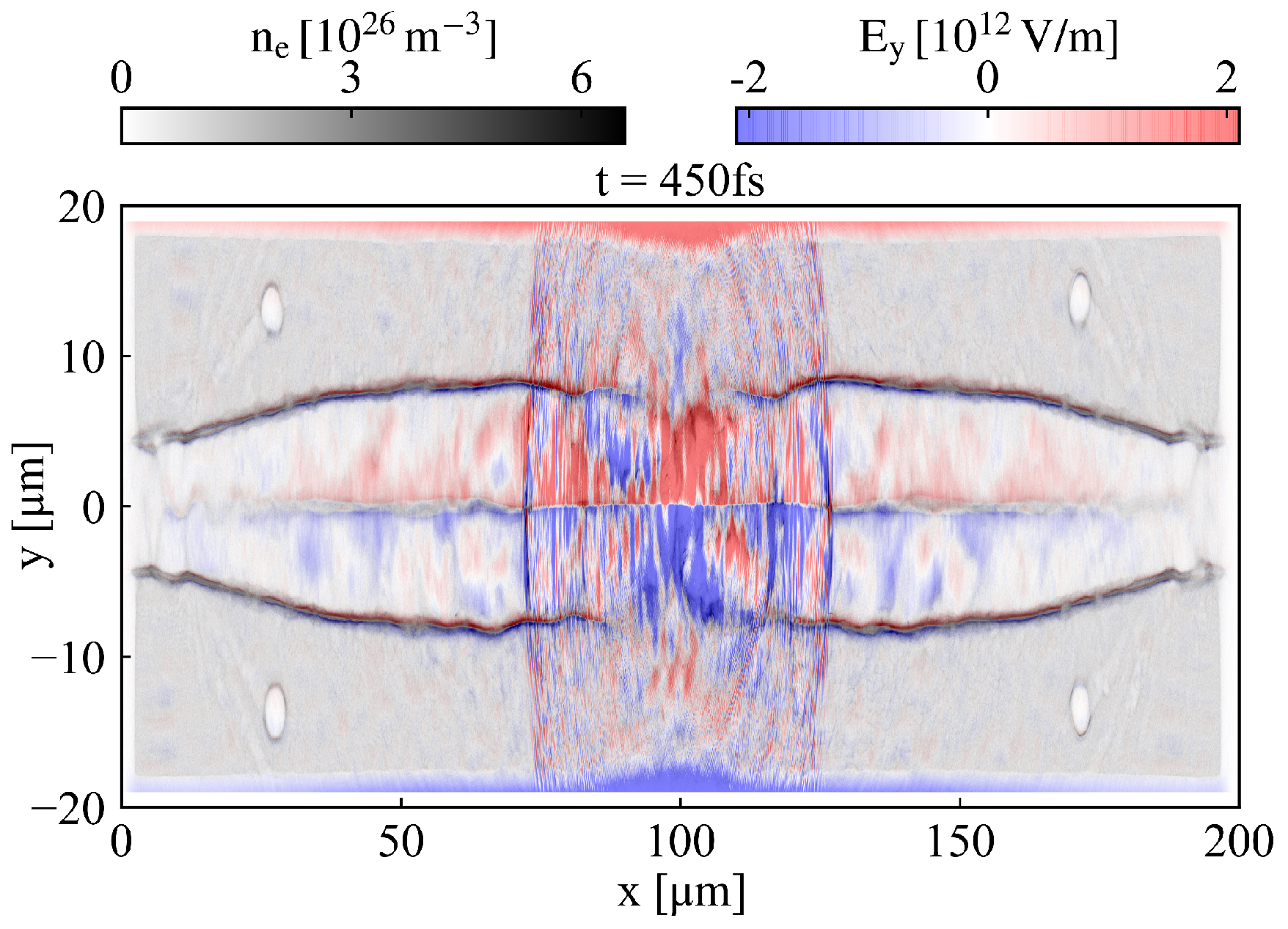}
    \caption{The electron number density distribution and transverse field at 450 fs from 2D-PIC the simulation of $\sim$200 $\mu$m SCDP driven by the laser pulse of case A. }
    \label{fig6}
\end{figure}

Towards the demonstration of photon-photon collision in the laboratory, the prerequisite is the ability to detect positrons in the experiments \cite{yu2019}. This section investigates the potential diagnostic methods by exploring the dynamics of positrons in a 2D-PIC simulation of $\sim$200 $\mu$m SCDP driven by the laser pulses of case A. The simulation reveals that a strong transverse electric field will be generated during the collision. This field prompts a large number of electrons to move out of the collision area from the transverse direction, and suppresses the positrons moving away until $\sim$440 fs. Figure \ref{fig6} shows the electron number density distribution and transverse field at 450 fs. From this figure, one can see that there is a quasi-static transverse electric field of 10$^{12}$V/m in the plasma channel. This electric field could accelerate the positrons out of the collision area from the transverse direction.

\begin{figure}[htb]
    \centering
    \includegraphics[width=1\linewidth]{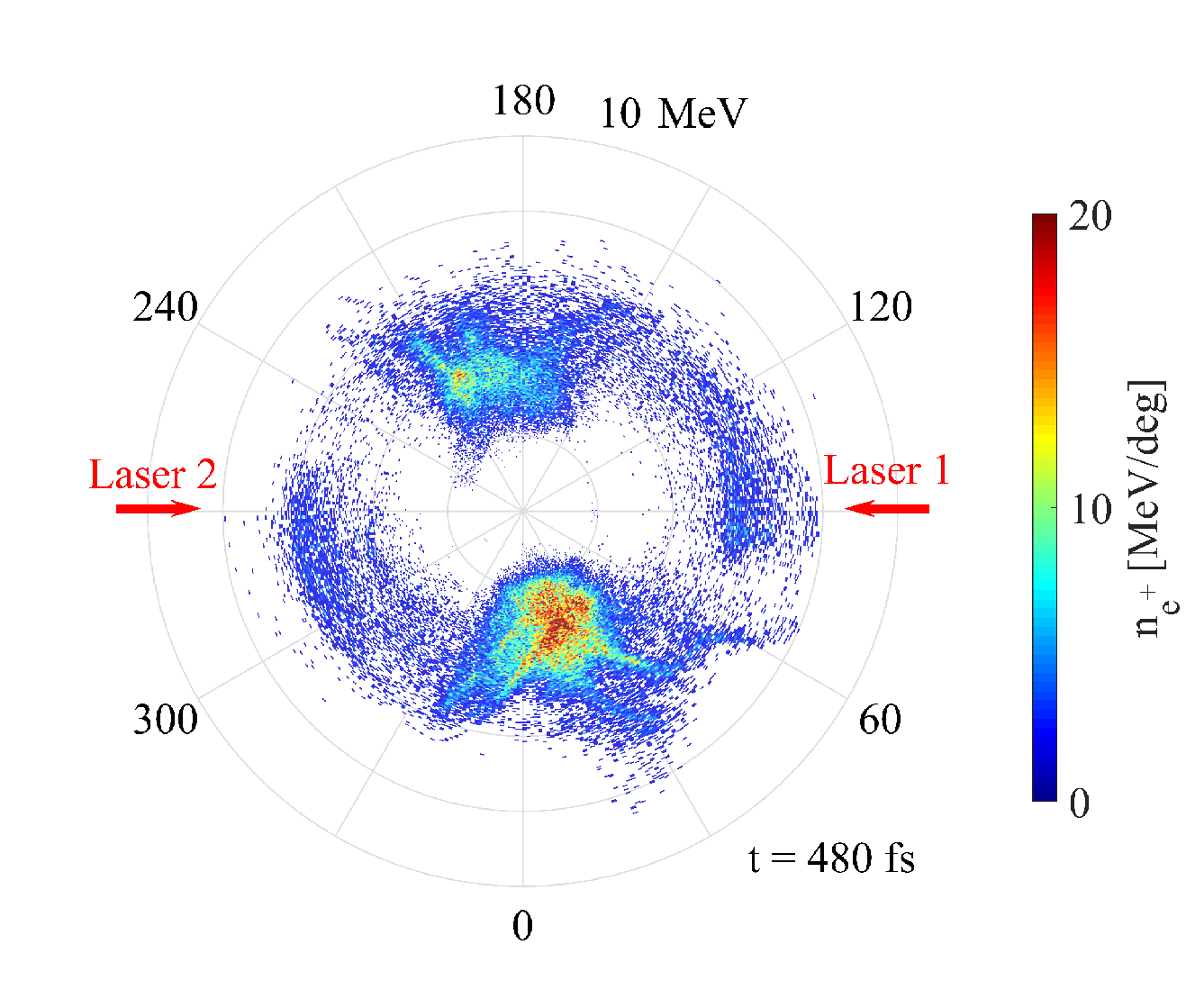}
    \caption{ The angular-spectrum distribution of the positrons at 480 fs. The positrons are accelerated transversely by the quasi-static transverse electric field. The central energy of the positrons is $\sim$3.4 MeV.}
    \label{fig7}
\end{figure}

Then, we inputted the full information of the created positrons into the simulation, and made the positrons to move self-consistently in the plasma starting from 400fs. Since the density of the positron is much smaller than that of electrons and ions, the impact of positrons on the fields can be ignored. From the simulation, it is found that the positrons begin to experience transverse acceleration from 440 fs, and most of the positrons could be accelerated off the plasma channel before $\sim$480 fs. Figure \ref{fig7} shows the angular-spectrum distribution of the positrons at 480 fs. The figure shows that most of the positrons could be accelerated to $\sim$3.4 MeV in directions divergent from the transmission direction of the laser pulses. Hence, this proposal could enable the safe detection of positrons in the direction away from the laser pulses.

\section{ Summary }

In summary, we propose a scheme to observe the two-photon Breit-Wheeler process in a single shot utilizing tightly focused 100s TW lasers which can be easily obtained on compact platforms. The interaction between the laser pluses and a sub-critical density plasma (SCDP) generates two high-charge electron bunches of 6 nC with an energy of 300 MeV. When the electron bunch collides with the laser pluse, Thomson scattering \cite{thomson1883,yan2017} takes place, resulting in the production of high brilliance X-ray pulses, whose photon number is higher than $3\times 10^{11}$, and brilliance is above $1.6\times 10^{23}$ photons/s/mm$^2$/mrad$^2$/0.1$\%$BW at 1 MeV. The X-ray pulses collide within a spot about 2 $\mu$m and create more than $1.1\times 10^5$ electron-positron pairs. The optimal plasma length is around 250 $\mu$m for a given $a_0$, while the photon number and pair number are proportional to $a_0^{2.4}$ and $a_0^{5.0}$ respectively. A 2D-PIC simulation demonstrates that the positrons can be accelerated by a strong transverse field and move away from the plasma channel. This scheme solves the key challenges of experimentally observing two-photon Breit-Wheeler process on compact platforms, particularly the limitation of generating electron-positron pairs in a single shot.

\begin{acknowledgments}
This work was supported by the Natural Science Foundation of China (Grant Nos. 11921006, 12175058), Beijing distinguished young scientist program and National Grand Instrument Project No. SQ2019YFF01014400. The PIC code EPOCH was in part funded by United Kingdom EPSRC Grant Nos. EP/G054950/1, EP/G056803/1, EP/G055165/1, and EP/M022463/1.
\end{acknowledgments}

\bibliography{apssamp}

\end{document}